\documentclass{article}
\usepackage{amsmath}
\usepackage{amsfonts}

\usepackage{amsmath,amssymb}
\usepackage{graphicx}
\usepackage{color}

\newtheorem{theorem}{Theorem}
\newtheorem{lemma}{Lemma}

\newtheorem{remark}{Remark}

\newtheorem{example}{Example}

\makeatletter

\newcommand{\Rmnum}[1]{\expandafter\@slowromancap\romannumeral #1@}
\makeatother

\newcommand{\done}{\hfill $\Box$ }

\newcommand{\ls}[1]
    {\dimen0=\fontdimen6\the\font\lineskip=#1\dimen0
     \advance\lineskip.5\fontdimen5\the\font
     \advance\lineskip-\dimen0
     \lineskiplimit=0.9\lineskip
     \baselineskip=\lineskip
     \advance\baselineskip\dimen0
     \normallineskip\lineskip\normallineskiplimit\lineskiplimit
     \normalbaselineskip\baselineskip
     \ignorespaces}

\begin{document}

\bibliographystyle{abbrv}

\title{$p$-ary sequences with six-valued cross-correlation
function: a new decimation of Niho type}
\author{Yuhua Sun$^{1}$, Hui Li$^{1}$, Zilong Wang$^{2}$, and Tongjiang Yan$^{3,4}$\\
$^1$ Key Laboratory of Computer Networks and
Information Security, \\
Xidian University,
Xi'an 710071, Shaanxi,
China\\
$^2$ State Key Laboratory of Integrated Service Networks,\\
Xidian University,
Xi'an, 710071, Shaanxi,
China\\
$^3$ College of Mathematics and Computational Science, \\
China University of Petroleum£¬
Dongying\ 257061,\ Shandong,\ China\\
$^4$ State Key Laboratory of Information Security (Institute of
Software,\\
 Chinese Academy of Sciences),
Beijing 100049,China\\
}

 \maketitle

\footnotetext[0] {The work is supported by the National Natural Science Foundation of China (No.60772136), Shandong Provincial Natural Science Foundation of China (No. ZR2010FM017), the Fundamental Research Funds for the Central Universities(No.K50510010012) and the open fund of State Key Laboratory of Information Security(Graduate University of Chinese Academy of Sciences)(No.F1008001).}

\thispagestyle{plain} \setcounter{page}{1}

\begin{abstract}

For an odd prime $p$ and $n=2m$, a new decimation $d=\frac{(p^{m}-1)^{2}}{2}+1$ of Niho type of $m$-sequences is presented. Using generalized Niho's Theorem, we show that the cross-correlation function between a $p$-ary $m$-sequence of period $p^{n}-1$ and its decimated sequence by the above $d$ is at most six-valued and we can easily know that the magnitude of the cross correlation is upper bounded by $4\sqrt{p^{n}}-1$.

\end{abstract}

\ls{1.5}
\section{Introduction}

It is of great interest to find a decimation value $d$ such that the cross-correlation between a $p$-ary $m$-sequence $\{s_{t}\}$ of period $p^{n}-1$ and its decimation $\{s_{dt}\}$ is low. For the case $\mathrm{gcd}(d,p^{n}-1)=1$, the decimated sequence $\{s_{dt}\}$ is also an $m$-sequence of period $p^{n}-1$. Basic results on the cross-correlation between two $m$-sequences can be found in \cite{Helleseth1}, \cite{Helleseth3}, \cite{Niho} and \cite{Trachtenberg}. For the case $\mathrm{gcd}(d,p^{n}-1)\neq1$, the reader is referred to \cite{Muller}, \cite{Hu}, and \cite{Choi}.

Let $p$ be a prime and $n=2m$. The cross correlation functions for the type of decimations $d\equiv1\ (\mathrm{mod}\ p^{m}-1)$ were first studied  by Niho \cite{Niho} named as Niho type of decimations. In \cite{Niho}, for $p=2$ and $d=s(p^{m}-1)+1$, Niho converted the problem of finding the values of cross-correlation functions into the problem of determining the number of solutions of a system of equations . This result is called Niho's Theorem. In 2006, Rosendahl \cite{Rosendahl} generalized Niho's Theorem to nonbinary sequences. In 2007, Helleseth et al. \cite{Helleseth2} proved that the cross correlation function between two $m$-sequences that differ by a decimation $d$ of Niho type is at least four-valued.

When $d=2p^{m}-1\equiv1\ (\mathrm{mod}\ p^{m}-1)$, the cross correlation function between a $p$-ary $m$-sequence $\{s_{t}\}$ of period $p^{2m}-1$ and its decimated sequence $\{s_{dt}\}$ is four-valued, which was originally given by Niho \cite{Niho} for the case $p=2$ and by Helleseth \cite{Helleseth1} for the case $p>2$. And when $d=3p^{m}-2$, the cross correlation function between two $m$-sequences that differ by $d$ is at most six-valued, especially, for $p=3$, the cross correlation function is at most five-valued \cite{Rosendahl}.

In this note, we study a new decimation
$d=\frac{(p^{m}-1)^{2}}{2}+1$ of Niho type. Employing generalized
Niho's Theorem, we show that the cross-correlation function between a $p$-ary
$m$-sequence and its decimated sequence by $d$ is at most
six-valued.

The rest of this note is organized as follows. Section 2 presents
some preliminaries and definitions. Using generalized Niho's
Theorem, we give an alternative proof of a result by Helleseth
\cite{Helleseth1} where $d=\frac{(p^{m}-1)(p^{m}+1)}{2}+1$ in section 3. A
new decimation $d=\frac{(p^{m}-1)^{2}}{2}+1$ of Niho type is given
in section 4. We prove that the cross correlation function between a
$p$-ary $m$-sequence and its decimated sequence by $d$ takes at most
six values.

\section{Preliminaries}

We will use the following notation in the rest of this note.
Let $p$ be an odd prime, $\mathrm{GF}(p^{n})$ the finite field with $p^{n}$ elements and $\mathrm{GF}(p^{n})^{*}=\mathrm{GF}(p^{n})\backslash\{0\}$.
The trace function $\mathrm{Tr}_{m}^{n}$ from the field $\mathrm{GF}(p^{n})$ onto the subfield $\mathrm{GF}(p^{m})$ is defined as
$$\mathrm{Tr}_{m}^{n}(x)=x+x^{p^{m}}+x^{p^{2m}}+\cdots+x^{p^{(l-1)m}},$$
where $l=\frac{n}{m}$ is an integer.

We may assume that a $p$-ary $m$-sequence $\{s_{t}\}$ of period $p^{n}-1$ is given by
$$s_{t}=\mathrm{Tr}_{1}^{n}(\alpha^{t}),$$
where $\alpha$ is a primitive element of the finite field $\mathrm{GF}(p^{n})$ and $\mathrm{Tr}_{1}^{n}$ is the trace function from $\mathrm{GF}(p^{n})$ onto $\mathrm{GF}(p)$.
The periodic cross correlation function $C_{d}(\tau)$ between$ \{s_{t}\}$ and its decimated sequence $\{s_{dt}\}$ is defined as
$$C_{d}(\tau)=\Sigma_{t=0}^{p^{n}-2}\omega^{s_{t+\tau}-s_{dt}},$$
where $\omega=e^{\frac{2\pi\sqrt{-1}}{p}}$ and $0\leq \tau \leq p^{n}-2$.

We will always assume that $n=2m$ is even in this note unless otherwise specified.

\section{An alternative proof of a known result}

For $p=2$,  Niho \cite{Niho} presented Niho's Theorem about decimations of Niho type of $m$-sequences. Rosendahl \cite{Rosendahl} generalized this result as follows.

\begin{lemma}
(generalized Niho's Theorem) \cite{Rosendahl} Let $p$, $n$,
and $m$ be defined as in section 2. Assume that $d\equiv1\
(\mathrm{mod}\ p^{m}-1)$, and denote $s=\frac{d-1}{p^{m}-1}$. Then
when $y={\alpha}^{\tau}$ runs through the nonzero elements of the
field $\mathrm{GF}(p^{n})$, $C_{d}(\tau)$ assumes exactly the values
$$-1+\left(N(y)-1\right)\cdot p^{m},$$
where $N(y)$ is the number of common solutions of
\begin{align}
x^{2s-1}+y^{p^{m}}x^{s}+yx^{s-1}+1&=0,\nonumber\\
x^{p^{m}+1}&=1.\nonumber
\end{align}
\end{lemma}

In 1976, Helleseth \cite{Helleseth1} proved the following result. Here, using the generalized Niho's Theorem, we give a simple proof.

\begin{theorem}
\cite{Helleseth1} Let the symbols be defined as in
section 2, $p$ be an odd prime and $d=\frac{p^{n}-1}{2}+1$. Then
$C_{d}(\tau)\in \{-1-p^{m}$, $-1$, $-1+p^{m}$,
$-1+\frac{p^{m}-1}{2}p^{m}, -1+\frac{p^{m}+1}{2}p^{m}\}$.
\end{theorem}
\vspace{0.1in} \noindent {\bf Proof of Theorem 1.}
Since $d=\frac{p^{n}-1}{2}+1=\frac{p^{m}+1}{2}(p^{m}-1)+1$, we get $s=\frac{d-1}{p^{m}-1}=\frac{p^{m}+1}{2}$. By Lemma 1, we have  $$C_{d}(\tau)=-1+\left(N(y)-1\right)\cdot p^{m},$$
where $y=\alpha^{\tau}$, $0\leq\tau\leq p^{n}-2$, and $N(y)$ is the number of common solutions of
\begin{align}
x^{(p^{m}+1)-1}+y^{p^{m}}x^{\frac{p^{m}+1}{2}}+yx^{\frac{p^{m}+1}{2}-1}+1&=0,\label{3.1}\\
x^{p^{m}+1}&=1.\label{3.2}
\end{align}
Note that Eq. (\ref{3.2}) implies
\begin{align}
x^{\frac{p^{m}+1}{2}}=1\label{3.3}
\end{align}
or
\begin{align}
x^{\frac{p^{m}+1}{2}}=-1.\label{3.4}
\end{align}
Substituting (\ref{3.3}) and (\ref{3.4}) into (\ref{3.1}) respectively, we get
$$C_{d}(\tau)=-1+\left(N_{1}(y)+N_{-1}(y)-1\right)\cdot p^{m},$$
where $N_{1}(y)$ is the number of the common solutions of
$$
(3.1.1)\ \ \left\{
\begin{array}{ll}
(y^{p^{m}}+1)x+(y+1)=0,\\
x^{\frac{p^{m}+1}{2}}=1,
\end{array}
\right.
$$
and $N_{-1}(y)$ is the number of solutions of
$$
(3.1.2)\ \ \left\{
\begin{array}{ll}
(y^{p^{m}}-1)x+(y-1)=0,\\
x^{\frac{p^{m}+1}{2}}=-1.
\end{array}
\right.
$$
Obviously, for $y\neq\pm1$, $0\leq N_{1}(y)+N_{-1}(y)\leq2.$

Let $y=1$. First, it is straightforward to get
$N_{-1}(1)=\frac{p^{m}+1}{2}$. Second, we see that $x=-1$ is the
only solution of (3.1.1) for $p^{m}+1\equiv0\ \mathrm{mod}\ 4$ and
$N_{1}(1)=0$ for $p^{m}+1\equiv2\ \mathrm{mod}\ 4$. Hence, we have
$$
N_{1}(1)+N_{-1}(1)=\left\{
\begin{array}{ll}
1+\frac{p^{m}+1}{2},\ \ \ \mathrm{if}\ p^{m}+1\equiv0\ \mathrm{mod}\ 4,\\
\frac{p^{m}+1}{2},\ \ \ \ \ \ \ \ \mathrm{if}\ p^{m}+1\equiv2\ \mathrm{mod}\ 4.
\end{array}
\right.
$$
Similarly, for $y=-1$, we have
$$
N_{1}(-1)+N_{-1}(-1)=\left\{
\begin{array}{ll}
\frac{p^{m}+1}{2},\ \ \ \ \ \ \ \ \mathrm{if}\ p^{m}+1\equiv0\ \mathrm{mod}\ 4,\\
1+\frac{p^{m}+1}{2},\ \ \ \mathrm{if}\ p^{m}+1\equiv2\ \mathrm{mod}\ 4.
\end{array}
\right.
$$
The result follows.\done

In Theorem 1, the value $s$ of Niho type decimation $d$ is equal to
$\frac{p^{m}+1}{2}$ corresponding to Lemma 1. Motivated by the above
proof, we take $s$ as the value $\frac{p^{m}-1}{2}$, a new
decimation of Niho type will be presented, and cross correlation
values will be determined in the following section.

\section{A new decimation of Niho type}
In this section, we give a new decimation $d$ of Niho's type and we show that the cross correlation function between a $p$-ary $m$-sequence and its decimated sequence by $d$ is at most six-valued.

\begin{theorem}
Let the symbols be defined as in section 2. Let $d=\frac{(p^{m}-1)^{2}}{2}+1$. Then $C_{d}(\tau)\in\{-1+(j-1)\cdot p^{m}|\ 0\leq j\leq5\}$ is at most six-valued.
\end{theorem}
\vspace{0.1in} \noindent {\bf Proof of Theorem 2.}
since
$d=\frac{(p^{m}-1)^{2}}{2}+1=\frac{p^{m}-1}{2}\cdot(p^{m}-1)+1\equiv1\ \mathrm{mod}\ (p^{m}-1),$
we know that the value $s$ corresponding to that in Lemma 1 is $\frac{p^{m}-1}{2}$. By the same argument as in Theorem 1, we get
$$ C_{d}(\tau)=-1+\left(N_{1}(y)+N_{-1}(y)-1\right)\cdot p^{m},$$
where $N_{1}(y)$ is the number of solutions of
$$
(4.1.1)\ \ \left\{
\begin{array}{ll}
x^{3}+y^{p^{m}}x^{2}+yx+1=0,\\
x^{\frac{p^{m}+1}{2}}=1,
\end{array}
\right.
$$
and $N_{-1}(y)$ is the number of solutions of
$$
(4.1.2)\ \ \left\{
\begin{array}{ll}
x^{3}-y^{p^{m}}x^{2}-yx+1=0,\\
x^{\frac{p^{m}+1}{2}}=-1.
\end{array}
\right.
$$
By the basic algebraic theorem, we know that $0\leq N_{1}(y)\leq3$ and $0\leq N_{-1}(y)\leq3$, i.e., $0\leq N_{1}(y)+N_{-1}(y)\leq 6$. Further, we will prove $0\leq N_{1}(y)+N_{-1}(y)\leq 5$, i.e., we will prove $N_{1}(y)+N_{-1}(y)\neq6$.

Suppose that $N_{1}(y)+N_{-1}(y)= 6$. Then $N_{1}(y)=3$ and $N_{-1}(y)=3$, i.e., both (4.1.1) and (4.1.2) have three solutions. Now, for $i=1,2,3$, let $x_{i}$ and $x_{i}^{\ast}$ be the solutions of (4.1.1) and (4.1.2), respectively. Since $x_{i}$ satisfies $x^{\frac{p^{m}+1}{2}}=1$ and $x_{i}^{\ast}$ satisfies $x^{\frac{p^{m}+1}{2}}=-1$, we know that there exists  some even integer $j_{i}$ satisfying $x_{i}=\alpha^{j_{i}(p^{m}-1)}$ and that there exists some odd integer $j_{i}^{\ast}$ satisfying $x_{i}^{\ast}=\alpha^{j_{i}^{\ast}(p^{m}-1)}$, where $i=1,2,3$. Simultaneously, since $x_{i}$ and $x_{i}^{\ast}$ satisfy the first equations of (4.1.1)and (4.1.2) respectively, we have
\begin{align}
\prod\limits_{i=1}^{3}x_{i}&=\alpha^{(p^{m}-1)\sum\limits_{i=1}^{3}j_{i}}=-1,\nonumber\\
\prod\limits_{i=1}^{3}x_{i}^{\ast}&=\alpha^{(p^{m}-1)\sum\limits_{i=1}^{3}j_{i}^{\ast}}=-1.\nonumber
\end{align}
By multiplying the above two equations, we get
$$\prod\limits_{i=1}^{3}x_{i}\prod\limits_{i=1}^{3}x_{i}^{\ast}=\alpha^{(p^{m}-1)(\sum\limits_{i=1}^{3}j_{i}+\sum\limits_{i=1}^{3}j_{i}^{\ast})}=1,$$
and induce
$p^{m}+1|\sum\limits_{i=1}^{3}j_{i}+\sum\limits_{i=1}^{3}j_{i}^{\ast}$.
This contradicts to the fact that $p^{m}+1$ is even but
$\sum\limits_{i=1}^{3}j_{i}+\sum\limits_{i=1}^{3}j_{i}^{\ast}$ is
odd. Therefore, we get $N_{1}(y)+N_{-1}(y)\neq6$, i.e., $0\leq
N_{1}(y)+N_{-1}(y)\leq 5$. The result follows.\done

\begin{remark}
The decimated sequence $\{s_{dt}\}$  in Theorem 2
is not necessarily an $m$-sequence. In fact,
$d=\frac{p^{m}-1}{2}(p^{m}-1)+1\equiv
\frac{p^{m}-1}{2}(-2)+1\equiv3\ \mathrm{mod}\ (p^{m}+1)$. For
$p\equiv-1\ \mathrm{mod}\ 3$, $m$ odd, we know that
$\mathrm{gcd}(d,p^{n}-1)=3$, $\{s_{dt}\}$ is not an $m$-sequence.
For the other case, $\mathrm{gcd}(d,p^{n}-1)=1$, and $\{s_{dt}\}$ is an $m$-sequence.
\end{remark}
\begin{remark}
Theoretically, the number of the values of $C_{d}(\tau)$ can not be reduced to less than 6.
Following is an example whose cross correlation function between an $m$-sequence and its decimated sequence by $d$ has exactly six values.
\end{remark}
\begin{example}
Let $p=3$, $n=6$, $m=3$ and $d=\frac{(p^{m}-1)^{2}}{2}+1=339$. The polynomial $f(x)=x^{6}+x^{5}+2$ is primitive over $\mathrm{GF}(3)$. Let $\alpha$ be a root of $f(x)$, then $s_{t}=\mathrm{Tr}_{1}^{6}(\alpha^{t})$, $s_{dt}=\mathrm{Tr}_{1}^{6}(\alpha^{339t})$. Computer experiment gives the following cross correlation values:
$$
\begin{array}{lclc}
-1-3^{3}&\mathrm{occurs}&246&\mathrm{times},\\
-1  &\mathrm{occurs}&284&\mathrm{times},\\
-1+3^{3}&\mathrm{occurs}&144&\mathrm{times},\\
-1+2\cdot 3^{3}&\mathrm{occurs}&42&\mathrm{times},\\
-1+3\cdot 3^{3}&\mathrm{occurs}&3&\mathrm{times},\\
-1+4\cdot 3^{3}&\mathrm{occurs}&9&\mathrm{times}.
\end{array}
$$
\end{example}

\section*{Conclusion}
In this note, using generalized Niho's Theorem, we give an
alternative proof of a known result. By changing the form of the
known decimation factor, we give a new decimation
$d=\frac{(p^{m}-1)^{2}}{2}+1$ of Niho type. We prove that the
cross correlation function between a  $p$-ary $m$-sequence of
period $p^{n}-1$ and its decimated sequence by the value $d$ is at
most six-valued, and we can easily see that the magnitude of the
cross correlation values is upper bounded by $4\sqrt{p^{n}}-1$.

\end{document}